\documentclass[sigconf]{acmart}

\usepackage{bm}
\usepackage{svg}
\usepackage{algorithm}
\usepackage{algpseudocode}
\usepackage{multicol}
\usepackage{multirow}
\usepackage{balance}

\usepackage{float}  
\usepackage{graphicx}

\AtBeginDocument{%
  \providecommand\BibTeX{{%
    \normalfont B\kern-0.5em{\scshape i\kern-0.25em b}\kern-0.8em\TeX}}}

\copyrightyear{2023}
\acmYear{2023}
\setcopyright{acmlicensed}\acmConference[KDD '23]{Proceedings of the 29th ACM SIGKDD Conference on Knowledge Discovery and Data Mining}{August 6--10, 2023}{Long Beach, CA, USA}
\acmBooktitle{Proceedings of the 29th ACM SIGKDD Conference on Knowledge Discovery and Data Mining (KDD '23), August 6--10, 2023, Long Beach, CA, USA}
\acmPrice{15.00}
\acmDOI{10.1145/3580305.3599364} \acmISBN{979-8-4007-0103-0/23/08}

\begin{document}

\def\pdfshellescape{1}
\title{Generative Flow Network for Listwise Recommendation}


\author{Shuchang Liu}
\affiliation{%
  \institution{Kuaishou Technology}
  \city{Beijing}
  \country{China}}
\email{liushuchang@kuaishou.com}

\author{Qingpeng Cai}
\affiliation{%
  \institution{Kuaishou Technology}
  \city{Beijing}
  \country{China}}
\email{caiqingpeng@kuaishou.com}

\author{Zhankui He}
\affiliation{%
  \institution{University of California, San Diego}
  \city{California}
  \country{USA}}
\email{zhh004@eng.ucsd.edu}

\author{Bowen Sun}
\affiliation{%
  \institution{Peking University}
  \city{Beijing}
  \country{China}}
\email{bwzdbml@gmail.com}

\author{Julian McAuley}
\affiliation{%
  \institution{University of California, San Diego}
  \city{California}
  \country{USA}}
\email{jmcauley@cs.ucsd.edu}

\author{Dong Zheng}
\affiliation{%
  \institution{Kuaishou Technology}
  \city{Beijing}
  \country{China}}
\email{zhengdong@kuaishou.com}

\author{Peng Jiang$^\dagger$}
\thanks{$\dagger$ \text{Corresponding author}}
\affiliation{%
  \institution{Kuaishou Technology}
  \city{Beijing}
  \country{China}}
\email{jiangpeng@kuaishou.com}

\author{Kun Gai}
\affiliation{%
  \institution{Unaffliated} 
  \city{Beijing}
  \country{China}}
\email{gai.kun@qq.com}

\renewcommand{\shortauthors}{Shuchang Liu et al.}


\begin{abstract}
Personalized recommender systems fulfill the daily demands of customers and boost online businesses.
The goal is to learn a policy that can generate a list of items that matches the user's demand or interest.
While most existing methods learn a pointwise scoring model that predicts the ranking score of each individual item, recent research shows that the listwise approach can further improve the recommendation quality by modeling the intra-list correlations of items that are exposed together.
This has motivated the recent list reranking and generative recommendation approaches that optimize the overall utility of the entire list.
However, it is challenging to explore the combinatorial space of list actions and existing methods that use cross-entropy loss may suffer from low diversity issues.
In this work, we aim to learn a policy that can generate sufficiently diverse item lists for users while maintaining high recommendation quality.
The proposed solution, GFN4Rec, is a generative method that takes the insight of the flow network to ensure the alignment between list generation probability and its reward.
The key advantages of our solution are the log scale reward matching loss that intrinsically improves the generation diversity and the autoregressive item selection model that captures the item mutual influences while capturing future reward of the list.
As validation of our method's effectiveness and its superior diversity during active exploration, we conduct experiments on simulated online environments as well as an offline evaluation framework for two real-world datasets.
\end{abstract}





\maketitle

\section{Introduction}

Recommender systems present a list of items upon each user's request to fulfill their personalized demand and interest.
And the quality of the recommended list directly impacts the user's experience and his/her satisfaction with the overall system.
Abundant literature has studied various supervised learning approaches ~\cite{cheng2016wide, guo2017deepfm, liu2022neural} that increase the model expressiveness to better capture the patterns in the complex user-recommender interactions.
While most existing methods adopt a pointwise or pairwise learning-to-rank paradigm that results in a model that separately scores each individual item for ranking, evidence ~\cite{burges2010ranknet} has shown that optimizing a listwise utility appears to be a superior option since it tends to make better use of the item's mutual influences in the list.
As an intuitive example, adding an item with high click probability may not always produce better list-wise performance, since other items in the list might be too similar causing competition.
In contrast, adding an item with low click probability may not always produce worse list performance, since it may emphasize or complement the neighboring items and make them more attractive.
Based on this motivation, the list-wise ranking approaches~\cite{burges2010ranknet, ai2018learning} and slate recommendation methods~\cite{jiang2018beyond, eugene2019slateq} have been proposed.

The key challenge of solving the list-wise recommendation problem is how to effectively and efficiently search the combinatorially large action space. Existing work could generally be categorized as either learning a list-wise evaluator~\cite{feng2021grn} or learning a list-wise generator~\cite{jiang2018beyond}.
The first approach uses the evaluator to approximate the list-wise utility function to guide the generation of lists.
However, this paradigm heavily depends on the accuracy of the evaluator which makes it less promising in recommendation tasks.
The latter approach belongs to the generative methods that can model the intra-list patterns and the list utility together in the generative process.
Its stochastic generation process could greatly improve the diversity but with a severe trade-off on the recommendation quality (we show evidence in section \ref{sec: online_learning}).
As another challenge of the list-wise recommendation problem, an item list typically aggregates the probability of exposing high-quality items during recommendation and is less likely to explore lists with slightly lower utility.
This is especially true for standard training with cross-entropy loss, as we will illustrate in section \ref{sec: relation_to_existing_methods}.

To solve the aforementioned challenges, we reformulate the goal into providing sufficiently diverse and high-quality recommendation lists.
Intuitively, sufficient recommendation diversity would expand the policy's knowledge of the action space and improves its efficiency in finding better recommendation.
On the other hand, we would also want to make sure that the diverse recommendations have a high quality so that the search of item list could become more reasonable and improves the exploration effectiveness on the action space.
Thus, in this work, we propose a generative approach based on a new flow-matching learning paradigm~\cite{bengio2021flow,pan2022generative,pan2023better} which is capable of generating diverse and accurate recommendations.
The key insights behind the proposed framework consist of a flow-matching loss that directly aligns the list generation probability with the list's utility in log scale, and an autoregressive item selection model that iteratively appends an item into the output list.
Specifically, the autoregressive item selection process is associated with a generation tree, each possible list corresponds to a root-to-leaf trajectory, and the generative model controls the probability flow on the tree graph.
By matching the list-wise probability flow with the utility, the resulting methods tend to align the log-likelihood of an item with log scale rewards (rather than aligning with the original reward as in cross-entropy), which gives a higher chance of exposure for items with slight lower rewards.
One challenge during the optimization of our method is that the large action space may induce extremely skewed probability distribution towards zero, so bias factors are introduced to control the scale of the probability aggregation and stabilize the learning of the generative model.

We summarize our contributions as follows:
\begin{itemize}
    \item We propose the GFN4Rec framework for the listwise recommendation problem and discuss its relationships with existing generative and reinforcement learning approaches.
    \item We build simulated online environments based on two real-world datasets and validate the superiority of GFN4Rec over strong list-wise recommendation methods when training and exploring online, and prove its ability to provide diverse recommendations with high quality.
    \item We conduct offline training and evaluation on the datasets as well to validate the consistent performance of GFN4Rec and the feasibility of the online environment.
\end{itemize}


\section{Background}\label{sec: background}

\subsection{Problem Formulation}\label{sec: problem_formulation}

We define a set of user $\mathcal{U}$ and a set of item $\mathcal{I}$.
Each recommendation request from a user $u\in\mathcal{U}$ consists of a set of profile features (e.g. user ID, gender), the most recent history of interactions, and a candidate set $\mathcal{C}$.
Note that a multi-stage recommendation process will have $\mathcal{C}\subset\mathcal{I}$ and $\mathcal{C} = \mathcal{I}$ only holds for a one-stage recommendation task.
Specifically, we denote the recommendation in the first case ($\mathcal{C}\subset\mathcal{I}$) as a \textit{re-ranking} scenario where an initial ranker exists, and denote that in the second case ($\mathcal{C} = \mathcal{I}$) as a \textit{ranking} scenario.

\textbf{Goal:} Then, the goal is to learn a policy $\pi( \mathcal{C},u;\theta)$ that selects an item list $\mathcal{O}\in\mathcal{C}^K$ for the given user request and maximizes the listwise reward $\mathcal{R}(u,O)$.

We assume a multi-behavior scenario where the user may provide different types of feedback (e.g. click, like, comment) for each item exposure.
Formally, we define the set of user behavior as $\mathcal{B}$, and $y_{u,i,b}$ as the user $u$'s response of item $i$ with respect to behavior $b\in\mathcal{B}$.
Then, for a given list $\mathcal{O}=\{a_1,\dots,a_K\}$, each item $a_i$ obtains a multi-behavior response $Y_{u,a_i}=[y_{u,a_i,b_1},\dots,y_{u,a_i,b_{|\mathcal{B}|}}]$, and the list-wise user response is:
\begin{equation}
    Y_{u,\mathcal{O}} = \left[\begin{matrix}
        y_{u,a_1,b_1} & \dots & y_{u,a_K,b_1}\\
        \vdots & \ddots & \vdots\\
        y_{u,a_1,b_{|\mathcal{B}|}} & \dots & y_{u,a_K,b_{|\mathcal{B}|}}
    \end{matrix}\right]
\end{equation}
For simplicity, we define the listwise reward as the average of item-wise reward $\mathcal{R}(u,O) = \frac{1}{K}\sum_{i\in\mathcal{O}}\mathcal{R}(u,i)$, where the item reward is calculated as the weighted sum of different positive user responses $\mathcal{R}(u,i) = \sum_b w_b y_{u,i,b}$.
Note that this reward metric is linearly separable by items and linearly separable by behaviors, which can accommodate efficient pointwise/pairwise training.
However, it does not reflect the mutual influences of items so independently improving the item-wise reward $w_b y_{u,i,b}$ of a single item on a single behavior does not necessarily improves the list-wise metric, since the rewards of other items in the list may drop as consequences.
We remind readers that there are more advanced reward function designs that aim to improve the overall reward~\cite{zou2019reinforcement,cai2023twostage} and we consider them as complementary to our solution.

\textbf{Online vs Offline:} Additionally, we assume the existence of the online learning loop (data $\rightarrow$ policy $\rightarrow$ data) where the observed new interactions between $\pi$ and the user environment continuously expand the training data during the optimization of $\pi$.
This indicates that the policy's exploration ability also determines the knowledge it will learn in the future, which in turn affects the recommendation performance.
Note that this is different from the standard reinforcement learning setting in recommendation~\cite{eugene2019slateq,afsar2022reinforcement,xie2022long,liu2023exploration} and conventional session-based recommendation~\cite{wang2021survey} where the recommender needs to consecutively interact with a user for several rounds (one recommendation list in each round) and optimize the multi-round cumulative reward.
In our setting, the aforementioned learning goal is a single-list reward optimization goal, and we want to achieve it in a dynamic online environment.

\subsection{Related Work}\label{sec: related_work}

\textbf{Top-K Recommendation and List-wise Recommendation}: Standard pointwise and pairwise learning-to-rank methods~\cite{koren2009matrix,rendle2010factorization,cheng2016wide,guo2017deepfm,kang2018self,sun2019bert4rec} aims to learn an item-wise scoring function for a given user request, so they can adopt efficient supervise learning (by formulating the problem as classification task) and their expressiveness mainly comes from the sophisticated design of user request encoder (e.g. DNN~\cite{cheng2016wide}, Transformer~\cite{kang2018self}).
During inference, items are ranked based on the learned pointwise scoring function, and the top K items are selected as the recommendation.
Yet, this learning paradigm does not align with real-world recommendation services which present to the user a list of items at a time.
In such cases, the way how items are organized also influences how users respond to each item.
For example, some users might prefer more diverse recommendations while other users might want to compete for similar items in the same list~\cite{kunaver2017diversity}.
Then, the list-wise recommendation problem is defined to emphasize the mutual influences between items in the exposed list~\cite{cao2007learning,xia2008listwise,burges2010ranknet,ai2018learning,pei2019personalized}.
The general idea is to infer and learn from the difference between the inclusion and exclusion of a certain item in the exposed list with respect to the list-wise metric (e.g. NDCG) or the whole list evaluation (for more sophisticated rewards).
Some work also shows that in a multi-stage recommendation system, the reranking model can better model the item correlations since the candidate set size is significantly reduced enabling a more powerful neural model~\cite{pei2019personalized,feng2021grn,liu2022neural}.

\textbf{Generative List Recommendation}: In recent years, there has been a discussion on the generative perspective of the pointwise recommendation~\cite{wang2017irgan,liang2018variational} listwise recommendation~\cite{jiang2018beyond,liu2021pivot} or slate recommendation~\cite{eugene2019slateq}. 
To handle the enormous combinatorial output space of lists, the generative approach models the distribution of recommended lists directly and generates a list as a whole with the use of deep generative models. 
For example, ListCVAE~\cite{jiang2018beyond} uses Conditional Variational Autoencoders (CVAE) to capture the item positional biases and item interdependencies in list distribution. 
Although promising, subsequent research~\cite{liu2021pivot} has shown that ListCVAE struggles with accuracy-diversity trade-offs. 
Such an analysis shows that balancing the exploitation and exploration in existing generative list recommendation models remains challenging. 
Our method also belongs to the generative approach, but it uses a brand new flow matching paradigm~\cite{bengio2021flow} that directly maps the list generation probability with its utility. 
This learning scheme has the potential to generate high-quality recommendations with sufficient significantly improved diversity, which helps the online exploration and searching for a better recommendation.

\subsection{Preliminary on GFlowNet}\label{sec: gfn_preliminary}

The idea of GFlowNet~\cite{bengio2021flow} aroused first in the problem of stochastic object generation from a sequence of actions. 
For example, constructing and designing a molecular graph for new medicine.
And the main insight behind GFlowNet is considering the iterative object generation sequence $\tau = \{\mathcal{O}^0 \rightarrow \mathcal{O}^1 \rightarrow \dots \rightarrow \mathcal{O}^T\}$ as a trajectory in a probabilistic flow network, and the learned generative model aims to assign each trajectory a sampling probability proportional to the corresponding reward of the completed object:
\begin{equation}
    P(\tau) \propto R(\mathcal{O}^T)\label{eq: gfn_prop_obj}
\end{equation} 
similar to the energy-based generative model~\cite{lecun2006ebm}.
In order to avoid an expensive MCMC process, the proposed method borrows the idea of temporal difference~\cite{sutton2018reinforcement} in reinforcement learning and formulates a flow matching objective $\forall \mathcal{O}^{t+1} \in \tau$ as in Eq.\eqref{eq: flow_matching_condition}.
It ensures that the sum of incoming flow matches the sum of outgoing flow.
The reward has $\mathcal{R}=0$ for intermediate nodes and $\mathcal{R}>0$ only on leaf nodes, and the transition function $T$ states a deterministic object transformation based on the given action.
\begin{equation}
    \sum_{\substack{\mathcal{O}^t,a^t: \\T(\mathcal{O}^t,a^t)=\mathcal{O}^{t+1}}} \mathcal{F}(\mathcal{O}^t,a^t) = \mathcal{F}(\mathcal{O}^{t+1}) = \mathcal{R}(\mathcal{O}^{t+1}) + \sum_{a^{t+1}} \mathcal{F}(\mathcal{O}^{t+1},a^{t+1})\label{eq: flow_matching_condition}
\end{equation}

The author further derived two variants of this objective that are easy to optimize~\cite{malkin2022trajectory}, namely, the Detailed Balance (DB) loss and the Trajectory Balance (TB) loss:
\begin{equation}
\begin{aligned}
    \mathcal{L}_\text{DB}(\mathcal{O}^t,\mathcal{O}^{t+1}) = \left(\log \frac{\mathcal{F}(\mathcal{O}^t)P(\mathcal{O}^{t+1}|\mathcal{O}^t;\theta)}{\mathcal{F}(\mathcal{O}^{t+1}) P_B(\mathcal{O}^t|\mathcal{O}^{t+1};\theta)}\right)^2\\
    \mathcal{L}_\text{TB}(\tau) = \left(\log \frac{ Z_\theta \prod_{t=1}^T P(\mathcal{O}^{t}|\mathcal{O}^{t-1};\theta)}{\mathcal{R}(\mathcal{O}^T)\prod_{t=1}^T P_B(\mathcal{O}^{t-1}|\mathcal{O}^{t};\theta)}\right)^2\label{eq: gfn_tb_and_db}
\end{aligned}
\end{equation}
which involves the learning of a flow estimator $\mathcal{F}(\mathcal{O})$, a forward probability function $P(\mathcal{O}^t| \mathcal{O}^{t-1})$ that serves as the step-wise stochastic policy that builds up the object, and a backward probability function $P_B(\mathcal{O}^{t-1}|\mathcal{O}^t)$ that helps infer the flow from a certain parent.
The TB loss minimizes the difference between the trajectory flow and the observed reward, and it reaches the minimum when the forward inference and the backward inference are identical.
The DB loss optimizes the flow matching objective for each generation step $\mathcal{O}^t \rightarrow \mathcal{O}^{t+1}$, and for the leaf node with no child node, the denominator is replaced by the reward $\mathcal{R}(\mathcal{O}^T)$

In our setting of list recommendation, we found two critical components of GFlowNet that are most helpful in improving recommendation performances: 
a) The log-scale reward that increases the chance of exploring diverse item lists during online learning; And
b) the auto-regressive generation that optimizes a future reward while capturing the mutual influences of items.
We will further explain this in the next section.

\section{Proposed Method}\label{sec: method}

In this section, we illustrate our proposed framework GFN4Rec.
Compared to GFlowNet's original design, our solution framework adopts several key changes to accommodate the list recommendation problem stated in section \ref{sec: problem_formulation}:
a) The generation of a recommendation list forms a tree graph rather than a directed acyclic graph, which means that the backward probability is always one;
b) The models are conditioned on user request $u$ so that collaborative learning can be used to alleviate the limited samples per request;
c) The action space (i.e. item list) is usually much larger than that in~\cite{bengio2021flow} indicating a harder exploration problem, so we add bias terms for the global normalization, the reward scaling, and the forward probability shift to stabilize the training.

\subsection{Item Selection Model and Generation Tree}\label{sec: generation_tree}

\begin{figure}[t]
    \centering
    \includegraphics[width=\linewidth]{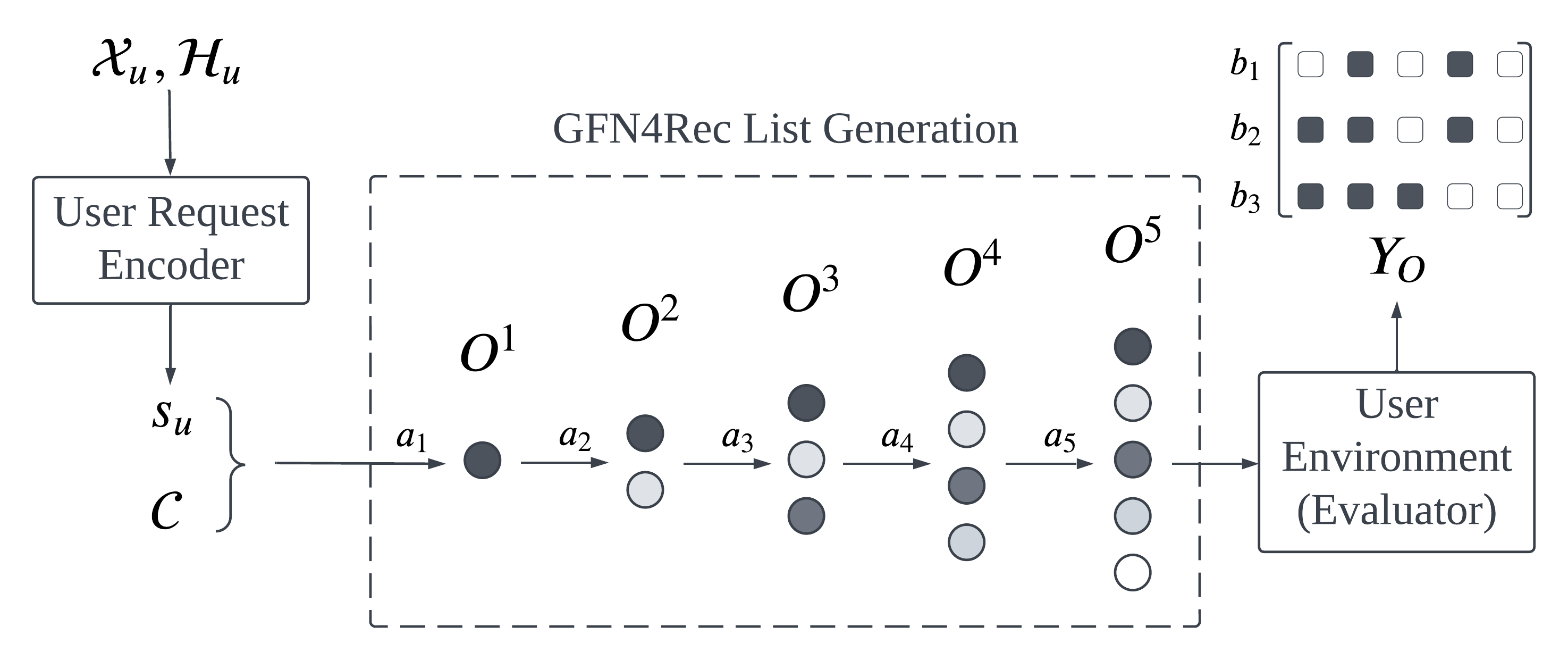}
    \caption{Example of list generation with $K=5$ and three types of user responses.}
    \label{fig: list_gen_mdp}
\end{figure}

We follow an autoregressive generation process that selects one item at a time.
During inference, a user request $u$ comes in and it contains the user information (profile features $\mathcal{X}_u$ and recent history $\mathcal{H}_u$), and the initial output list is empty, i.e. $\mathcal{O}^{0}=\varnothing$.
At each step $t>0$, an item $a_t\in\mathcal{C}/\mathcal{O}^{t-1}$ is selected based on the probabilistic model $a_t\sim P_\theta(i|u,\mathcal{O}^{t-1})$, noted as the \textit{item selection model}, parameterized by $\theta$.
Then the selected item is pushed at the end of the output list, i.e. $\mathcal{O}^{t} = \mathcal{O}^{t-1}\oplus\{a_t\}$ is an ordered list.
At the final step $t=K$, we will have a full recommendation list $\mathcal{O}^K = \{a_1, \dots, a_K\}$ which is then exposed to the user environment in answer to the request.
Figure \ref{fig: list_gen_mdp} shows an example of this process with $K=5$ and the item selection model in each step is presented in Figure \ref{fig: gfn_model}.
During online exploration, the item is randomly sampled based on the softmax score, and for greedy strategies, we select the item with the top score.
Note that our problem focuses on the list-wise recommendation, and there is no intermediate response signal for an item selection step until the final list is generated.

\begin{figure}[t]
    \centering
    \includegraphics[width=0.9\linewidth]{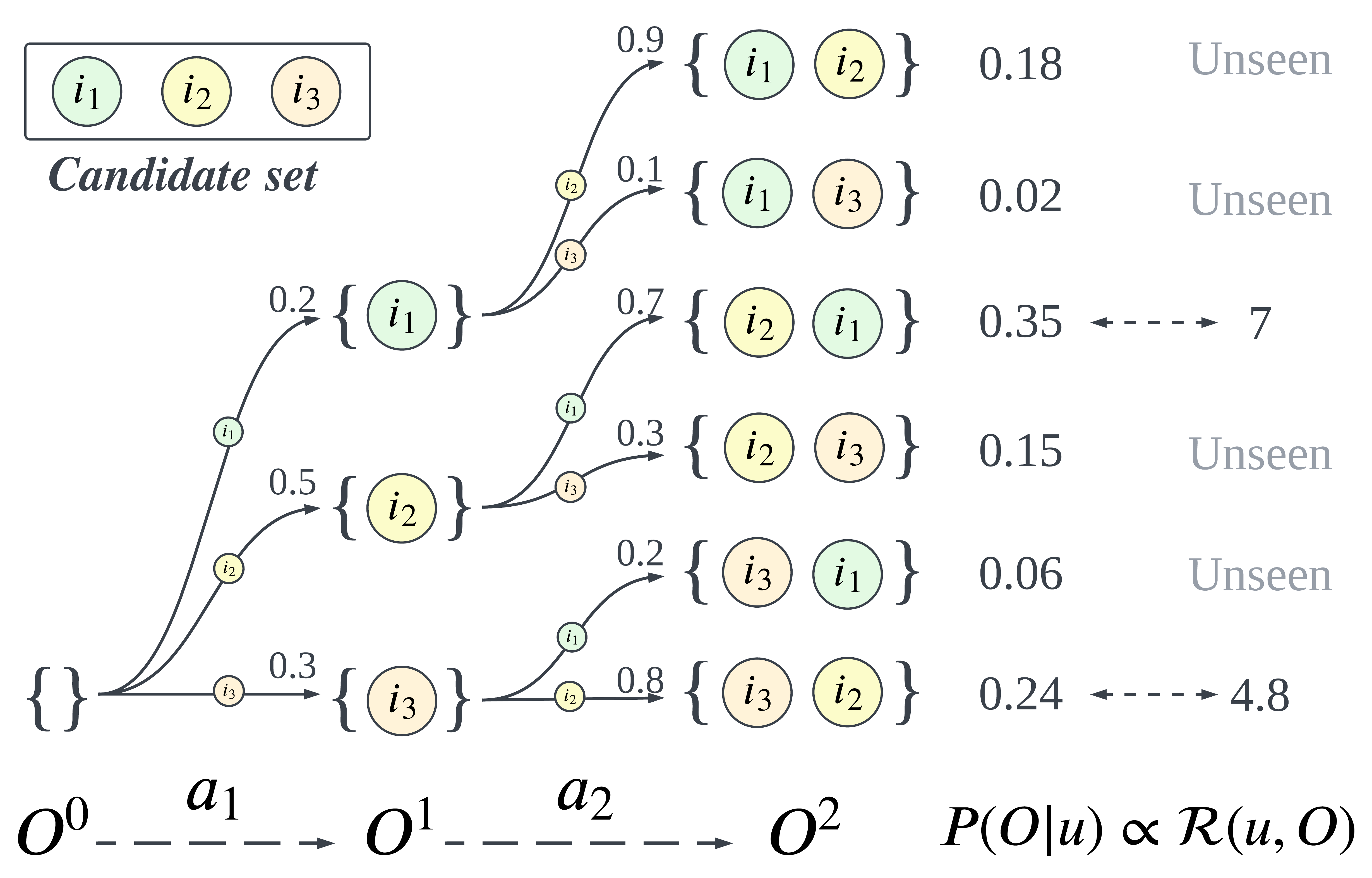}
    \caption{Example of generation tree with K=2, $|\mathcal{C}|=3$.}
    \label{fig: gen_tree_example}
\end{figure}

\textbf{The generation tree:} We assume a recommendation list of a fixed size $K$ (also known as the slate recommendation). 
Since we iteratively add items into the list in order, the generation graph of all possible lists forms a $K$-depth tree structure, where the nodes are (intermediate or final) output lists and each edge represents a selected item.
Figure \ref{fig: gen_tree_example} shows an example of such a generation tree.
In a tree graph, each node $\mathcal{O}^t$ has only one possible parent node $\mathcal{O}^{t-1}$ except for the source node that has no parent.
And the number of children for a given node $\mathcal{O}^t$ is linear to $|\mathcal{C}|-t$ except the leaf nodes that have no child.
All leaves have depth $K$, and the total number of leaves (i.e. list-wise search space) is equivalent to the number of size-$K$ placement: ${|\mathcal{C}| \choose K}\times K! = O(|\mathcal{C}|^K)$.
By sampling according to the autoregressive item selection model $P(a_t|u,\mathcal{O}^{t-1})$, the generator ends up with a trajectory with the observed output list $\mathcal{O}=\mathcal{O}^K=\{a_1, \dots, a_K\}$, and the output list (in leaf node) has a one-to-one correspondence to its generation trajectory.
Thus, we can obtain the generation probability of the output list as its unique trajectory's sampling probability conditioned on $u$:
\begin{equation*}
    P(\mathcal{O}|u) = \prod_{t=1}^{K} P(\mathcal{O}^t|u,\mathcal{O}^{t-1}) = \prod_{t=1}^{K} P_\theta(a_t|u,\mathcal{O}^{t-1})
\end{equation*}
where the choice of item $a_t$ determines the output list in the next step, i.e. $P(\mathcal{O}^t|u,\mathcal{O}^{t-1}) = P_\theta(a_t|u,\mathcal{O}^{t-1})$.
Using the example in Figure \ref{fig: gen_tree_example}, the recommendation $\{i_2, i_1\}$ has a trajectory probability $P(i_2|u,\varnothing)P(i_1|u,\{i_2\}) = 0.5 \times 0.7 = 0.35$.

\subsection{Learning Objectives on Network Flow}\label{sec: method_learning_objectives}

Different from the standard reward maximization goal in most learning-to-rank paradigms, we want to learn a generative model that not only finds the best reward but also favors other high-reward recommendation lists for better exploration.
Thus, following Eq.\eqref{eq: gfn_prop_obj}, we aim to learn a trajectory distribution that is proportional to the list-wise rewards for a certain user $u$:
\begin{equation}
    P(\mathcal{O}|u) \propto \mathcal{R}(u,\mathcal{O})\label{eq: gfn_goal}
\end{equation}
As we will discuss in section \ref{sec: relation_to_existing_methods}, this would enforce the model to match the log scale rewards for items that are less likely to be trapped in local sub-optima and boosts the exploration of lists with slightly lower rewards.
One challenge of the optimization under this learning goal is the limited observation per user request (or only one interaction per request in the most extreme case).
Fortunately, we can solve this through collaborative training across users.

\textbf{Matching the flow and the reward: } Intuitively, users have different behavioral patterns which induce different reward distributions.
In order to match these differences, we assign a personalized initial flow estimator $\mathcal{F}(u,\mathcal{O}^0)=\mathcal{F}(u,\varnothing)$ to the source node (the starting step with an empty list), representing the prior of the reward.
Then the generation tree will split this initial flow according to the step-wise item selection model and the flow of a leaf node with $\mathcal{O}$ is $\mathcal{F}(u,\mathcal{O}) = \mathcal{F}(u,\varnothing) P(\mathcal{O}|u)$.
Combining with Eq.\eqref{eq: gfn_goal}, the user-wise flow distribution will have:
\begin{equation}
    b_z\mathcal{F}(u,\mathcal{O}) = \mathcal{R}(u,\mathcal{O})
\end{equation}
where $b_z$ is a hyperparameter that represents the fixed global normalizing bias for the forward passes compared to observed rewards.

\textbf{Learning the trajectory probability:} Based on previous notions, for an observed training sample $(u, \mathcal{O}, \mathcal{R}(u,\mathcal{O}))$, we can derive from Eq.\eqref{eq: gfn_tb_and_db} the trajectory balance (TB) objective:
\begin{equation}
    \mathcal{L}_\mathrm{TB} = \Big(\log b_z + \log \frac{\mathcal{F}_\phi(u,\varnothing)\prod_{t-1}^K P_\theta(a_t|u,\mathcal{O}^{t-1})}{\mathcal{R}(u,\mathcal{O}^K) + b_r}\Big)^2
\end{equation}
where $b_r$ is a hyperparameter that represents the global reward bias, and it is introduced to control the smoothness of the loss landscape and avoids division by zero rewards.
The learnable parameters include $\phi$ of the initial flow estimator $\mathcal{F}$ and $\theta$ of the item selection model (representing the forward probability function).
Note that the backward probability is a constant $P(\mathcal{O}^{t-1}|u, \mathcal{O}^t)=1$ since each node has only one parent in a tree graph.

\textbf{From trajectory-wise to step-wise: } the TB loss optimizes the overall trajectory as a whole but induces a large variance in the squared error term.
One alternative is to use a more detailed objective (derived from the DB loss of Eq.\eqref{eq: gfn_tb_and_db}) on each item generation step $\mathcal{O}^{t-1}\rightarrow\mathcal{O}^{t}$:
\begin{equation}
    \mathcal{L}_\mathrm{DB} = \begin{cases}
        \Big(\log\mathcal{F}_\phi(u,\mathcal{O}^K) - \log(\mathcal{R}(u,\mathcal{O}^K)+b_r)\Big)^2 & \text{for leaf node}\\
        \Big(\frac{\log b_z}{K} + \log \frac{\mathcal{F}_\phi(u,\mathcal{O}^{t-1})P_\theta(a_t|u,\mathcal{O}^{t-1})}{\mathcal{F}_\phi(u,\mathcal{O}^t)P(\mathcal{O}^{t-1}|u, \mathcal{O}^t)}\Big)^2 & t\in\{1,\dots,K\}
    \end{cases}
\end{equation}
It consists of a reward-matching term for the leaf node and a flow-matching term for each of the intermediate nodes.
Here, $\mathcal{F}_\phi(\cdot)$ represents the flow estimator for any given node (leaf or intermediate), and the reward smooth bias $b_r$ and normalizing bias $b_z$ have the same meaning as in $\mathcal{L}_\mathrm{TB}$.
Again, the single-parent property of nodes in a tree graph gives $P(\mathcal{O}^{t-1}|u, \mathcal{O}^t)=1$ and we can simplify the second case of $\mathcal{L}_\mathrm{DB}$ to:
\begin{equation}
    \mathcal{L}_\mathrm{DB} = \Big(\frac{\log b_z}{K} + \log \frac{\mathcal{F}_\phi(u,\mathcal{O}^{t-1})P_\theta(a_t|u,\mathcal{O}^{t-1})}{\mathcal{F}_\phi(u,\mathcal{O}^t)}\Big)^2, t\in\{1,\dots,K\}
\end{equation}
Note that this learning objective is separable by item which is better suited for parallel training, but it does not directly optimize the trajectory probability, which may be less effective for limited observations or insufficient reward accuracy.

\textbf{Forward probability shifting for better stability} During training, we observe that the scale of $P(a_t|u,\mathcal{O}^{t-1})$ is usually around $\frac{1}{|\mathcal{I}|}$ which is quite different from the scale of the reward and the learned scale of the flow estimator.
This could induce a very large negative value with high variance after taking the log, which could dominate the gradient calculation at the beginning and makes the training process very unstable.
As a result, we also include a hyperparameter $b_f$ that shifts the forward probability to a value range similar to other components.
In other words, the original log term $\log P_\theta(a_t|u,\mathcal{O}^{t-1})$ is shifted to $\log (P(a_t|u,\mathcal{O}^{t-1}) + b_f)$.
As an intuitive example, we can set $b_f=1.0$ to make $\log (P(\cdot) + b_f)\geq 0$.

\subsection{Transformer-based User Request Encoder}\label{sec: user_request_encoder}

In our recommendation setting, a user request consists of the user's profile $\mathcal{X}_u$ that maintains the static features of the user as well as the L most recent interaction history $\mathcal{H}_u = [(a_1, Y_{a_1}),\dots,(a_L, Y_{a_L})]$ that captures the dynamic changes in the user's interest.
The user request encoder will take $\mathcal{X}_u$ and $\mathcal{H}_u$ as input and outputs a user state embedding $\bm{s}_u$ for later list generation phase.
It consists of a transformer-based history encoder and a DNN-based feature extraction module.
We present its details in Appendix \ref{app: model_detail}.
And we remind the readers that this request encoder is not specifically designed for our GFN4Rec method and it could accommodate many existing models that require a user encoding module~\cite{kang2018self} including the baselines in our experiments as described in section \ref{sec: experiments}.

\begin{figure}[t]
    \centering
    \includegraphics[width=\linewidth]{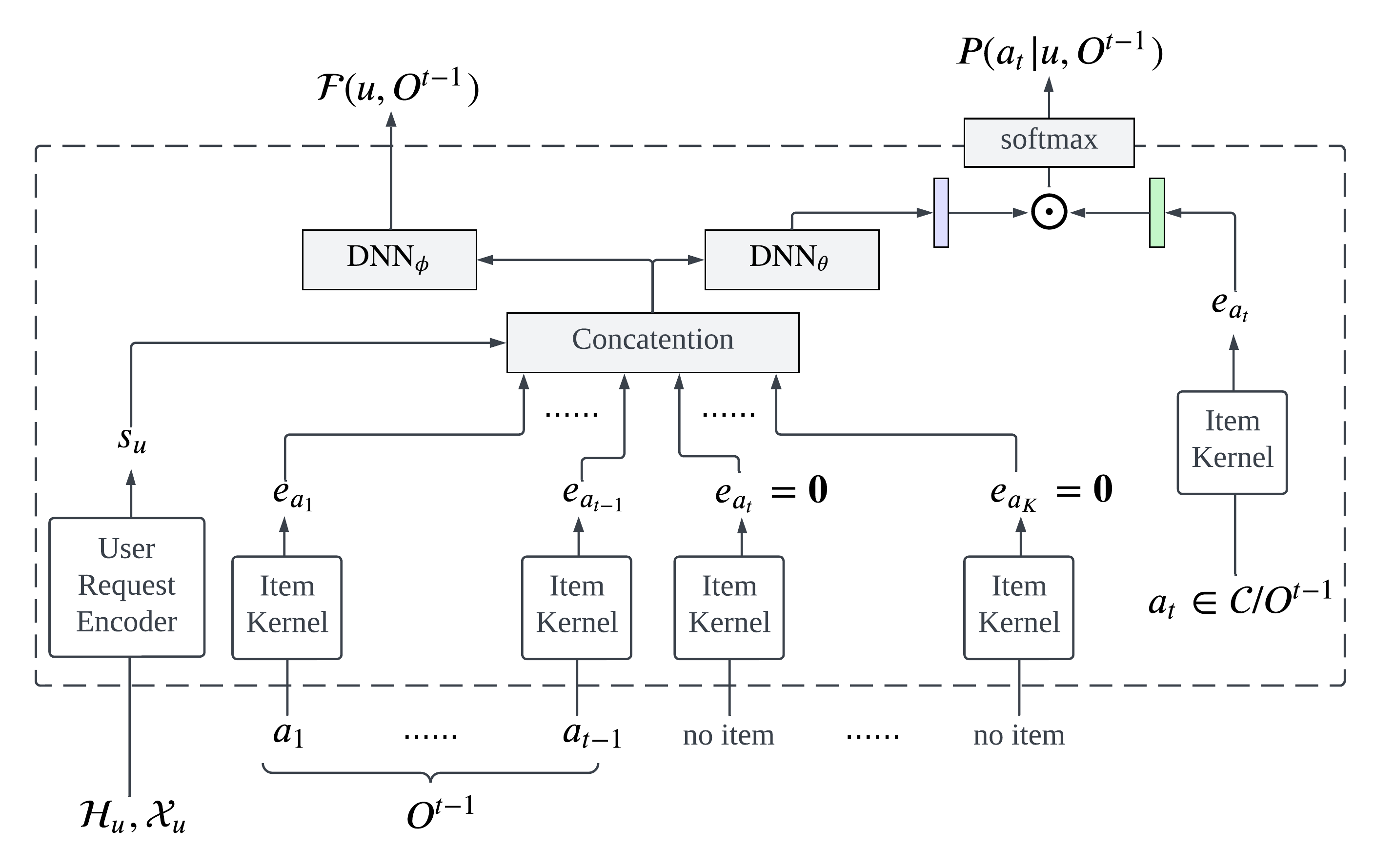}
    \caption{Flow estimator $\phi$ and item selection model $\theta$ in GFN4Rec. We presents details of the user request encoder and item kernel in Appendix \ref{app: model_detail}. $\odot$ represents dot product.}
    \label{fig: gfn_model}
\end{figure}

\subsection{Relation to Existing Methods}\label{sec: relation_to_existing_methods}

\textbf{Reward vs. Log-scale Reward:} In standard learning-to-rank solutions and many list-wise methods that assumes conditional independence of item probabilities, a classification paradigm is adopted, such as binary or multi-class cross-entropy loss \cite{rendle2009bpr, kang2018self, pei2019personalized}.
It results in an alignment between the item-wise log probability $P(i|u)$ and the item-wise reward, i.e. $\log P(i|u) \rightarrow \mathcal{R}(u,i)$.
Assuming independent item selection, then this would induce exponential probability aggregation for an item list: $P(\mathcal{O}^K|u) = \prod_{a_t\in\mathcal{O}^K} P(a_t|u) \rightarrow \prod_{a_t\in\mathcal{O}^K} e^{R(u,a_t)}$, which is sensitive to items with high scores.
Thus, the generator may quickly distinguish items with top-ranking scores and quickly converge to a local optimum.
In contrast, one of the key insights from the GFlowNet is the log scale reward matching paradigm, which aims to directly align the log probability with log-scaled reward, i.e. $\log P(\mathcal{O}|u) \rightarrow \log \mathcal{R}$.
Adopting the definition of list-wise reward in section \ref{sec: problem_formulation}, this log-scale alignment means that the list generation probability will be linear to the linear combination of item-wise reward: $P(\mathcal{O}^K|u) \rightarrow \mathcal{R}(u,\mathcal{O}) = \sum_{a_t\in\mathcal{O}}\mathcal{R}(u,a_t)$.
In such a case, items with high scores are less distinguishable than those with lower scores, and items with slightly lower point-wise scores now have a good chance of being selected.

\textbf{Evaluator vs. Generator:} As we have discussed in section \ref{sec: related_work}, list-wise recommendation approaches can be generally categorized as evaluator-based methods, generator-based methods, and evaluator-generator paradigms.
Our GFN4Rec framework is defined as a list generator where the list generation probability is proportional to its reward label.
Notably, this property also means that GFN4Rec can be regarded as an evaluator-based method as well since the trajectory probability $P(\mathcal{O}|u)$ estimated upon generation is also an indicator of the list's quality (represented by the list-wise reward).
This is different from generative methods like CVAE \cite{jiang2018beyond} that use the reward label as input upon generation.
In general, GFN4Rec as well as any generation model that matches the list generation probability with the reward is simultaneously a generator and an evaluator.
Compared to the generator-evaluator learning paradigm~\cite{feng2021grn} that uses a list evaluator to indirectly guide the recommendation policy, GFN4Rec is a more direct approach that is easier to optimize and stabilize.
Additionally, the autoregressive generation process of GFN4Rec does not restrict the model design and can accommodate many existing advanced solutions~\cite{bello2018seq2slate,sun2019bert4rec}, but the main difference lies in the flow matching loss for the entire list rather than learning from a decomposed item-wise signal.

\section{Experiments}\label{sec: experiments}

To validate the effectiveness of the proposed method, we conduct both offline and online experiments on two real-world datasets.

\textbf{Datasets:} we include two real-world datasets \textbf{ML1M} and \textbf{KR1K}.
ML1M is the one million version of the MovieLens dataset \footnote{https://grouplens.org/datasets/movielens/1m/} dataset that consists of users' rating (original range in $\{1,\dots,5\}$) history for movies, but the rating signals are transformed into clicks (rating $\geq$ 3), likes (rating $\geq$ 4), and stars (rating $\geq$ 5).
The KR1K is the 1K-user version of the KuaiRand~\cite{gao2022kuairand} dataset that consists of users' interaction histories for short videos, the user feedback include clicks, views, likes, comments, forwards, follows, and hates, and all behavior types are 0/1 signals \footnote{https://kuairand.com/}.
For both datasets, we filter the records into 20-core data and cut the user history into segments of size $K=6$, and regard each segment as an observed recommendation list.
For simplicity, we set the item-wise reward weight $w_b=1$ except that the hate signal in KR1K has $w_\mathrm{hate}=-1$.
As a result, the range of item-wise reward $\mathcal{R}(u,i)\in[0,3]$ in ML1M and $\mathcal{R}(u,i)\in[-1,6]$ in KR1K.
Statistics of the resulting datasets are summarized in Table \ref{tab: datasets}.

\begin{table}[t]
    \centering
    \begin{tabular}{c|ccccc}
        \hline
        Dataset & $|\mathcal{U}|$ & $|\mathcal{I}|$ & \#record & $|\mathcal{B}|$ & Range of $\mathcal{R}$ \\
        \hline
        ML1M & 6400 & 3706 & 1,000,208 & 3 & [0,3]\\
        KR1K & 1000 & 69,219 & 2,597,865 & 7 & [-1,6]\\
        \hline
    \end{tabular}
    \caption{Dataset Summary. The records are used for offline training of policies and online user environment, but not used for online training of policies.}
    \label{tab: datasets}
\end{table}

\textbf{Models and Baselines:}
We compare the GFN4Rec model with both ranking and reranking models. 
We summarize the included models as the following:
\begin{itemize}
    \item CF~\cite{kang2018self}: a pointwise model that scores the user-item interaction based on the dot product between the user encoding and the item encoding.
    \item ListCVAE ~\cite{jiang2018beyond}: a generative model that captures the list distribution based on conditional VAE, and the reward is formulated as the input condition when providing a recommendation.
    \item PRM~\cite{pei2019personalized}: a re-ranking model that uses the CF model as the initial ranker and uses a transformer-based re-ranker to encode the intermediate candidate set.
    \item GFN4Rec: our proposed GFN4Rec model with \textit{trajectory balance} loss. Comparison between trajectory balance and detailed balance will be further discussed in section \ref{sec: ablation}.
\end{itemize}
As mentioned in section \ref{sec: problem_formulation}, the ranking models provide a one-stage recommendation with $\mathcal{C}=\mathcal{I}$, and the re-ranking model is associated with a pretrained initial ranker that filters the item pool into a smaller candidate set $C\subset \mathcal{I}$ for the re-ranker.
To better control the variables in the model comparison, we use the same user request encoder across all models.
We present more model details in Appendix \ref{app: model_detail}.

\textbf{Simulated User Environment: } in order to simulate the complex multi-type user behavior in the observed data, we build a stochastic user response model $\mathcal{E}:\mathcal{U}\times\mathcal{C}^K\rightarrow \mathcal{B}^K$ that predict the probability of a user $u$ positively engage with item $i$ by behavior $b$.
The base neural model $g(u,\mathcal{O})$ outputs the initial behavior likelihood, and it consist of a Transformer-based user history encoder similar to the user request encoder, and a state-to-behavior predictor that infers the user response probability for the given recommendation $\mathcal{O}$.
We train this base model using binary cross entropy on the ground truth label $y_{u,i,b}$ and obtain AUC in $[0.7,0.9]$ for both datasets across different behaviors.
When the pretrained user response model takes effect in the online environment, we also include an item-influence module that suppresses the initial ranking score by each item's similarity to other items in the list, to simulate the user's demand for recommendation diversity.
We use a significance factor $\rho > 0$ to ensure the existence of item influence and set $\rho=0.2$ for ML1M while $\rho=0.1$ for KR1K.
The final user response $y_{u,i,b}$ is uniformly sampled based on the modified behavior likelihood to simulate the uncertainty of user feedback in the recommendation.
For the data sampling strategy of all online learning methods (e.g. In GFN4Rec, Algorithm \ref{alg: solution}, line 18), half of the mini-batch samples are newly added instances from the online inference procedure, and the other half comes from the uniform sampling over the entire buffer to avoid catastrophic forgetting~\cite{parisi2019continual}.

\begin{algorithm}[t]
\caption{GFN4Rec}
\begin{algorithmic}[1]
\Statex \# Apply current policy in running episodes:
\Procedure{Online Inference}{}{}
\State Initialize replay buffer $\mathcal{A}$.
\While{True, in each running episode}
    \State Observe user request $u$.
    \State Initial $\mathcal{O}^0\leftarrow \varnothing$
    \For{$t\in\{1,\dots,K\}$}
        \State Sample item $a_t\sim P_\theta(i|u,\mathcal{O}^{t-1})$ with current policy.
        \State $\mathcal{O}^{t} = \mathcal{O}^{t-1}\oplus\{a_t\}$
    \EndFor
    \State Obtain user responses $Y_\mathcal{O}$ from online environment and calculate $\mathcal{R}(u,\mathcal{O})$.
    \State $(u,\mathcal{O},\mathcal{R}(u,\mathcal{O}), Y_{u,\mathcal{O}})\rightarrow \mathcal{A}$
\EndWhile
\EndProcedure
\Statex \# Simultaneous training on the buffer:
\Procedure{Training}{}{}
\State Initialize all trainable parameters in the policy (e.g. $\theta$ and $\phi$ in GFN4Rec)
\State Wait until $\mathcal{A}$ has stored minimum amount of data points.
\While{Not Converged, in each iteration}
    \State Obtain mini-batch sample $(u,\mathcal{O},R(u,\mathcal{O}), Y_{u,\mathcal{O}}) \sim \mathcal{A}$.
    \State Calculate $P_\theta(a_t|u,\mathcal{O}^{t-1})$ and $\mathcal{F}_\phi(\mathcal{O}^t)$ for each generation step $t$.
    \State Update the policy through one step of gradient descent on $\mathcal{L}_\mathrm{TB}$ or $\mathcal{L}_\mathrm{DB}$.
\EndWhile
\EndProcedure
\end{algorithmic}\label{alg: solution}
\end{algorithm}

\begin{table*}[t]
    \small
    \centering
    \begin{tabular}{c|cccc|cccc}
        \hline
        \multirow{2}*{Method} & \multicolumn{4}{c}{ML1M} & \multicolumn{4}{c}{KR1K} \\
         & Avg. $\mathcal{R}$ & Max $\mathcal{R}$ & Coverage & ILD & Avg. $\mathcal{R}$ & Max $\mathcal{R}$ & Coverage & ILD \\
        \hline
        CF & 2.073 & 2.939 & 13.963 & 0.529 & \underline{2.253} & 4.039 & 100.969 & 0.543 \\
        ListCVAE & 0.940 & 2.209 & \underline{\textbf{262.420}} & \underline{\textbf{0.796}} & 2.075 & \underline{4.042} & \underline{\textbf{446.100}} & \underline{\textbf{0.565}} \\
        PRM & \underline{2.156} & \underline{2.967} & 18.647 & 0.559 & 2.174 & 3.811 & 27.520 & 0.538 \\
        \hline
        GFN4Rec(Explore) & 2.047 & 2.938 & 87.660 & 0.617 & 2.212 & 3.984 & 415.515 & \textbf{0.591} \\
        GFN4Rec & \textbf{2.172} & \textbf{2.972} & 15.693 & 0.565 & \textbf{2.414} & \textbf{4.054} & 21.267 &  0.520 \\
        \hline
    \end{tabular}
    \caption{Model performances of online learning model. Best values are in bold. Strongest baseline in underline.}
    \label{tab: online_results}
\end{table*}

\begin{table*}[t]
    \small
    \centering
    \begin{tabular}{c|cccccc|cccccc}
        \hline
        \multirow{2}*{Method} & \multicolumn{6}{c|}{ML1M} & \multicolumn{6}{c}{KR1K} \\
         & Avg. $\mathcal{R}$ & Max $\mathcal{R}$ & R-NDCG & R-MRR & Coverage & ILD & Avg. $\mathcal{R}$ & Max $\mathcal{R}$ & R-NDCG & R-MRR & Coverage & ILD \\
        \hline
        CF & 1.675 & 2.694 & 0.563 & 0.0713 & 12.217 & \underline{\textbf{0.729}} & \underline{1.941} & 3.860 & 0.390 & \underline{0.0824} & 17.275 & 0.611 \\
        ListCVAE & - & - & - & - & - & - & 1.896 & 3.802 & 0.381 & 0.0803 & \underline{\textbf{343.067}} & \underline{\textbf{0.657}} \\
        RerankCF & 1.901 & \underline{\textbf{2.918}} & 0.632 & 0.0806 & \underline{\textbf{129.823}} & 0.627 & 1.931 & \underline{\textbf{3.990}} & \underline{\textbf{0.395}} & 0.0823 & 153.186 & 0.586  \\
        PRM & \underline{1.914} & 2.914 & \underline{0.636} & \underline{0.0812} & 128.626 & 0.623 & 1.909 & \underline{3.966} & 0.386 & 0.0808 & 284.000 & 0.595\\
        \hline
        GFN4Rec & \textbf{1.996} & 2.908 & \textbf{0.665} & \textbf{0.0848} & 21.788 & 0.605 & \textbf{1.962} & 3.870 & 0.393 & \textbf{0.0834} & 32.16 & 0.630 \\
        \hline
    \end{tabular}
    \caption{Online simulator performances for offline model. Best values are in bold. Strongest baseline in underline. R-NDCG and R-MRR correspond to the R-NDCG(online) and R-MRR(online) metrics.}
    \label{tab: offline_results}
\end{table*}

\begin{table*}[t]
    \small
    \centering
    \begin{tabular}{c|cccc|cccc}
        \hline
        \multirow{2}*{Method} & \multicolumn{4}{c|}{ML1M} & \multicolumn{4}{c}{KR1K} \\
         & R-NDCG(online) & R-MRR(online) & R-NDCG(test) & R-MRR(test) & R-NDCG(online) & R-MRR(online) & R-NDCG(test) & R-MRR(test) \\
        \hline
        CF & 0.563 & 0.0713 & 0.533 & 0.0824 & 0.390 & 0.0824 & 0.356 & \underline{\textbf{0.0420}} \\
        ListCVAE & - & - & - & - & 0.381 & 0.0803 & \underline{\textbf{0.361}} & 0.0419 \\
        RerankCF & 0.632 & 0.0806 & 0.570 & 0.0835 & \underline{\textbf{0.395}} & \underline{0.0823} & 0.339 & 0.0415 \\
        PRM & \underline{0.636} & \underline{0.0812} & \underline{\textbf{0.578}} & \underline{\textbf{0.0861}} & 0.386 & 0.0808 & 0.352 & 0.0415\\
        \hline
        GFN4Rec & \textbf{0.665} & \textbf{0.0848} & 0.561 & 0.0826 & 0.393 & \textbf{0.0834} & \textbf{0.362} & \textbf{0.0421} \\
        \hline
    \end{tabular}
    \caption{Online and offline ranking metrics of offline model. Best values are in bold. Strongest baseline in underline.}
    \label{tab: ranking_metric}
\end{table*} 

\subsection{Online Learning}\label{sec: online_learning}

The main purpose of the online learning experiment is to 1) verify the GFN4Rec's ability to find better recommendation policies that produce higher rewards; 2) validate the more diverse behaviors of GFN4Rec during online sampling while keeping high-quality recommendations.

\subsubsection{Training framework: } we summarize the training procedures of GFN4Rec in algorithm \ref{alg: solution}.
Lines 18-20 correspond to the main optimization step and lines 5-9 are the online sampling steps.
During test time, if we aim to find the best output, the action sampling (in line 7) will be turned off and we will adopt greedy selection according to the scores provided by the item selection model.
To better illustrate the exploration behavior of our GFN4Rec method, we observe both the test performance under the aforementioned greedy selection and that using sampling (with line 7 turned on), we denote the latter as GFN4Rec(Explore).
When training other baselines, the overall online learning framework is similar to algorithm \ref{alg: solution} and differs mainly in the loss minimization step (lines 18-20) and the list generation step (lines 5-9).
For example, the CF baseline learns a pointwise model $P(i|u)$ which uses the dot product between user request encoding and candidate item kernel encoding as the ranking scores and simply selects the top-$K$ as the recommendation, and its objective function is the reward-based binary cross-entropy:
\begin{equation}
     \mathcal{L}_\mathrm{BCE} = - \mathcal{R}(u,i) \log P(i|u) + (1-\mathcal{R}(u,i)) \log (1-P(i|u))
\end{equation}
where the label in the original BCE loss is replaced by the continuous multi-behavior reward.
During training, we fix all experiments with a mini-batch size of 128 and start training after 100 steps of running episodes.
For reranking models, we include additional online training steps for the initial ranker before the training of the reranker, its learning objective also uses the aforementioned R-BCE loss.

\subsubsection{Evaluation Protocol: } For each user request and the recommended list, the online user environment returns the user feedback and we calculate the corresponding listwise reward $\mathcal{R}(u,\mathcal{O})$ (defined in section \ref{sec: problem_formulation}).
We report both the \textbf{Average Reward} as well as the \textbf{Max reward} across user requests in a mini-batch.
For diversity metrics, we include the item \textbf{Coverage} metric that describes the number of distinct items exposed in a mini-batch, and intra-list diversity (\textbf{ILD}) that estimates the embedding-based dissimilarity between items in each recommended list:
\begin{equation}
    \mathrm{ILD}(\mathcal{O}) = \frac{1}{K(K-1)}\sum_{a_i\in\mathcal{O}}\sum_{a_j\in \mathcal{O}/\{a_i\}} (1 - \mathrm{similarity}(a_i,a_j))
\end{equation}
As mentioned in~\cite{liu2021pivot}, the item coverage reflects the cross-list diversity which will help us understand how GFN4Rec generates diverse lists.
For each model, we use grid search to find the hyperparameters that yield the best results.
Specifically, we check learning rate in \{0.001, 0.0001, 0.00001\}, L2 regularization in \{0.0001, 0.00001, 0\}.
For ListCVAE, we search the $\beta$ coefficient of the KLD loss in \{1.0, 0.1, 0.01, 0.001\}.
For PRM, we control its PV loss coefficient in \{1.0, 0.1, 0.01\}.
For all GFN4Rec variants we search $b_r$ in \{0.1, 0.3, 1.0, 1.5\}, $b_f$ in \{0.1, 0.5, 1.0, 1.5, 2.0\}, and $b_z$ in \{0.1, 0.5, 1.0, 1.5\}.
We notice that most models converge around episode step 5000 in both ML1M and KR1K, and the average result of the last 100 steps is regarded as test samples for evaluation.

\subsubsection{Empirical Results: } 
After searching the model parameters, we run each model's best setting for 5 rounds with different random seeds and report the average test results in Table \ref{tab: online_results}.
In both online environments, GFN4Rec achieves the best performance in terms of the reward metrics, and it significantly outperforms the strongest baseline in KR1K by 10\% in the average reward.
The reranking PRM achieves the same level of reward in ML1M, but it takes advantage of an extra ranking phase.
This means that the GFN4Rec model can find a better recommendation policy than other baselines.
The online-sampling counterpart GFN4Rec(Explore) also achieves a relatively high reward (the same level as CF) in both environments, but what makes it superior is the significantly improved item coverage and ILD.
Specifically, in both ML1M and KR1K, GFN4Rec(Explore) improves the item coverage by 4$\times$ compared to CF and PRM.
ListCVAE could achieve the same level of diversity but suffers from severe accuracy trade-offs, especially in ML1M.
On the contrary, GFN4Rec(Explore) achieves almost the same level of diversity as ListCVAE, with a significantly better accuracy performance in terms of rewards.
All this evidence proves that GFN4Rec is able to find high-quality recommendations (in terms avg. $\mathcal{R}$ and max $\mathcal{R}$) with better diversity as an online learning framework.

\subsection{Offline Learning}\label{sec: offline_learning}

We include the offline experiments as verification of 1) the consistent performance of GFN4Rec in both offline and online evaluation; and 2) the feasibility of the online simulator (discussed in section \ref{sec: simulator_feasibility}).

\subsubsection{Training Framework: } For offline training, the policy no longer samples the lists online nor collects training samples into the buffer, so GFN4Rec(Explore) is no longer applicable.
Instead, it only uses the offline log data (as those in Table \ref{tab: datasets} that takes the same format $(u,\mathcal{O},\mathcal{R}(u,\mathcal{O}), Y_{u,\mathcal{O}})$.
Except for this difference in the data iterator, the remaining optimization steps are identical to the training procedure of algorithm \ref{alg: solution}.
To engage in offline test, we split the last $N$ interactions of each user's history as test samples while the remaining as training samples, and we set $N=1$ for ML1M and $N=4$ for KR1K.
We train each model with a mini-batch size of 128 and stop the training upon convergence (around 10000 steps in ML1M and 5000 steps in KR1K).
We exclude ListCVAE in the comparison of ML1M for its unstable and incomparable performance.

\subsubsection{Evaluation Protocol: } During the evaluation, for the data points in the test set, we modify the standard offline metrics into the reward-based NDCG (R-NDCG) and the reward-weighted mean reciprocal rank (R-MRR) as illustrated in Eq.\eqref{eq: metric}.
where the Rank$(u,i)$ is a position-wise rank of items on the same position in the batch data since each position in the recommendation list now corresponds to an item selection step.
The R-NDCG metric generalizes the standard NDCG metric where the item-wise reward $\mathcal{R}(u,i)$ becomes the relevance label, and the IDCG is agnostic to the model being evaluated.
The R-MRR metric generalizes the standard MRR metric but replaces the item label with the item-wise reward.
For both metrics, a larger value means that the learned policy performs better on the offline data.

\begin{equation}
\begin{aligned}
    \text{R-NDCG} & = \frac{1}{K} \sum_{k\in\{1,\dots,K\}} \text{R-NDCG}(k)\\ & 
    \text{R-NDCG}(k) = \frac{\sum_{u,a_k}\mathcal{R}(u,a_k) 2^{1-\mathrm{Rank}(u,a_k)}}{\mathrm{IDCG}}\\
    \text{R-MRR} & = \frac{1}{K} \sum_{k\in\{1,\dots,K\}} \text{R-MRR}(k)\\
    & \text{R-MRR}(k) = \sum_{u,a_k} \frac{\mathcal{R}(u,a_k)}{\mathrm{Rank}(u,a_k)}\\
    \label{eq: metric}
\end{aligned}
\end{equation}

Additionally, we can still deploy the models to the online environment even though they are trained offline, only that there is no buffer to maintain and no online sampling for exploration.
We adopt the same online evaluation protocol in section \ref{sec: online_learning} and include both the accuracy metrics (\textbf{average reward} and \textbf{maximum reward}) and the diversity metrics (item \textbf{Coverage} and \textbf{ILD}).
Note that we can calculate R-NDCG and R-MRR for both the offline data and the online observed interactions, so we denote the first case as R-NDCG(test) and R-MRR(test), and denote the second case as R-NDCG(online) and R-MRR(online).

\subsubsection{Empirical Results: } We adopt the same grid search for common hyperparameters as in online learning, and report the best parameter with 5-seed averaged results in Table \ref{tab: offline_results} and Table \ref{tab: ranking_metric}.
Specifically, in Table \ref{tab: ranking_metric}, GFN4Rec achieves better results than CF in ML1M and achieves the best results in KR1K in terms of the test set ranking metric R-NDCG(test) and R-MRR(test).
These offline metrics are almost consistent with the online metrics, but with one exception when comparing the reranking baseline PRM, where GFN4Rec is slighted better on R-NDCG(online) and R-MRR(online) and PRM is slightly better on R-NDCG(test) and R-MRR(test) in ML1M.
This might be related to the smaller action space of ML1M, which may improve the chance of the reranking mechanism to finding better intermediate candidates for later reranking.
In general, GFN4Rec is effective in finding better rewards than one-stage models when engaging in offline training, and its performance is consistent in both offline and online metrics.
Additionally, in Table \ref{tab: offline_results}, online ranking metrics (R-NDCG(online) and R-MRR(online)) are consistent with other online accuracy metrics (closest to Avg. $\mathcal{R}$) in terms of model comparison.
Combining with the aforementioned consistency between online and offline ranking metrics, this further verifies the feasibility of the evaluation under the online simulator (further explained in section \ref{sec: simulator_feasibility}).

\subsection{Ablation Study}\label{sec: ablation}

\begin{table}[t]
    \small
    \centering
    \begin{tabular}{c|cc}
        \hline
        \multirow{2}*{Method} & \multicolumn{2}{c}{KR1K} \\
         & GFN\_DB & GFN\_TB \\
        \hline
        Avg. $\mathcal{R}$ & \textbf{2.034} & 1.962 \\
        Max $\mathcal{R}$ & \textbf{3.905} & 3.870 \\
        Coverage & \textbf{2.034} & 1.962 \\
        ILD & 0.582 & \textbf{0.630} \\
        R-NDCG(online) & \textbf{0.400} & 0.393 \\
        R-MRR(online) & \textbf{0.0859} & 0.0834 \\
        R-NDCG(test) & \textbf{0.363} & 0.362 \\
        R-MRR(test) & \textbf{0.0423} & 0.0421 \\
        \hline
    \end{tabular}
    \caption{TB vs. DB with offline model training.}
    \label{tab: tb_vs_db_offline}
\end{table} 

\subsubsection{Trajectory Balance vs. Detailed Balance}\label{sec: tb_vs_db}

As we have discussed in section \ref{sec: method_learning_objectives}, trajectory balance $\mathcal{L}_\mathrm{TB}$ (denote as \textbf{GFN\_TB}) directly optimizes the item selection probability of different positions together, while the detailed balance $\mathcal{L}_\mathrm{DB}$ (denote as \textbf{GFN\_DB}) separates the learning of each position and only the last step is directly guided by the accurate reward label.
Thus, DB loss adopts step-wise learning which would result in lower variance in the squared error term, compared with TB loss.
This indicates that DB is potentially more suitable for larger action space (item candidate set).
As shown in Table \ref{tab: tb_vs_db_offline}, GFN\_DB achieves better performance than GFN\_TB in the practical KR1K dataset.
We suggest using DB loss in practice as it is more suitable for large action spaces and more stable in training.

\subsubsection{Greedy vs. Exploration}

\begin{figure}[t]
    \centering
    \includegraphics[width=\linewidth]{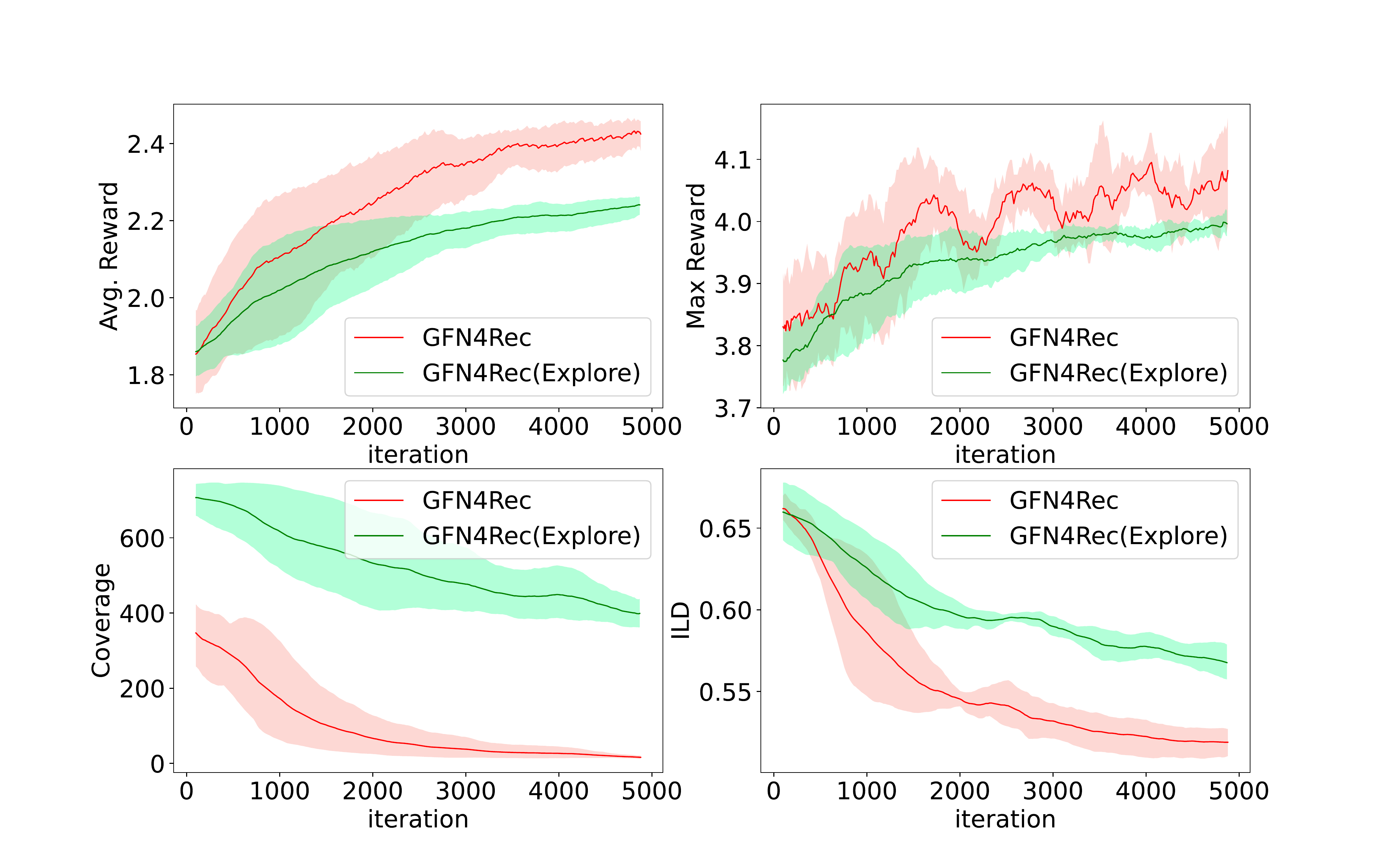}
    \caption{Learning curves of greedy GFN4Rec and GFN4Rec(Explore) in KR1K.}
    \label{fig: tb_vs_tbexplore}
\end{figure}

As supported by section \ref{sec: online_learning}, the GFN4Rec model can achieve high recommendation quality with better diversity than the exploration counterpart.
We further illustrate this in Figure \ref{fig: tb_vs_tbexplore}, where the reward metrics of GFN4Rec(Explore) grow much slower than that of the greedy GFN4Rec (for both DB and TB variants).
In contrast, the item coverage and ILD metrics drop much slower in GFN4Rec(Explore).
Additionally, we observe that the max reward, though it generally improves over time, appears to be very unstable.
GFN4Rec(Explore) exhibits very stable behavior, which indicates that there might exist a large number of slates with high quality while extreme actions could misguide the learning process.

\subsubsection{Feasibility of Online Simulator: }\label{sec: simulator_feasibility}

While offline metrics like NDCG and MRR are widely verified in practice, the feasibility of an online simulator for the recommendation has been an important research topic in recent years~\cite{ie2019recsim}.
We need a realistic online simulator that follows real-world user behavioral patterns in order to verify the effectiveness of recommendation policies.
In section \ref{sec: offline_learning}, we use both the online simulator and offline test set for model evaluation and observe that the two sets of metrics are generally consistent across different models.
This indicates that our design of the user environment is sufficiently realistic to model the user behaviors and feasible in validating the recommendation models.
Still, we remind readers that in theory, there is a feasible region of the pretrained user environment that is close to the offline data, but it does not exclude the chance of mode collapse if we do not regulate the pretraining process~\cite{luke2017unroll}.

\subsubsection{Offline vs. Online} 

As many online learning methods pointed out~\cite{xie2022long}, the offline log data does not provide the ground truth user feedback for the counterfactual question ``What if the policy recommends a different list and how would the user behave''.
This limitation restricts the exploration of better data samples and is the main motivation of the aforementioned research on user simulation.
In our experiments, we observe evidence of the sub-optimal reward performances in models that are trained offline compared with their online training counterparts.
For example, the best model in KR1K is GFN4Rec which achieves 2.414 in average reward, but it only reaches 1.962 on the same online simulator when it is trained offline.
This phenomenon is consistent across all variants of GFN4Rec, indicating the effectiveness of engaging exploration in the online environment and the limitation of offline training.

\subsubsection{Inconsistent Diversity of Reranking Model: } 

We observe that the reranking baseline PRM achieves significantly higher item coverage when trained with offline data but not so in online learning.
We believe this is related to the diversity of the initial ranker.
To validate this, we include the RerankCF baseline which consists of a CF-based initial ranker and a deterministic top-K selection as the re-ranker, and present its results in Table \ref{tab: offline_results}.
We observe that the diversity of RerankCF also achieves significantly higher item coverage than CF and GFN4Rec.
This indicates that the existence of the initial ranker could potentially improve the diversity but at a cost of lower accuracy (in online reward and offline metrics).

\section{Conclusion}\label{sec: conclusion}

In this work, we propose a list-wise recommendation method that uses a generative flow network to represent the probabilistic list generation process.
The resulting method GFN4Rec can generate high-quality recommendations with better diversity, which is well suited for the online learning environment that requires the guidance of exploration.
One key insight of the proposed method is a generative approach that directly matches the generation probability rather than the log probability with the list-wise reward.
Another feature of GFN4Rec is the iterative item generation model that captures the item's mutual information and optimizes a future list reward.
This notion may also suit other scenarios where intermediate rewards are not observed until the final object is generated (e.g. multi-stage recommendation).
However, we remind readers that GFN4Rec requires more hyperparameters (reward smoothing, normalizing coefficient, and forward probability offset) which may take empirical efforts to find a feasible optimization setting than standard supervised learning.
Additionally, GFN4Rec controls the balance between recommendation quality and diversity during online exploration, and the effectiveness of the exploration depends on the validity of this trade-off.
In general, it is still an open problem whether there exists a way to optimize the exploration strategy.

\bibliographystyle{ACM-Reference-Format}

\balance
\bibliography{main}

\appendix

\section{More Experimental Detail}

\begin{figure*}[t]
    \centering
    \includegraphics[width=\textwidth]{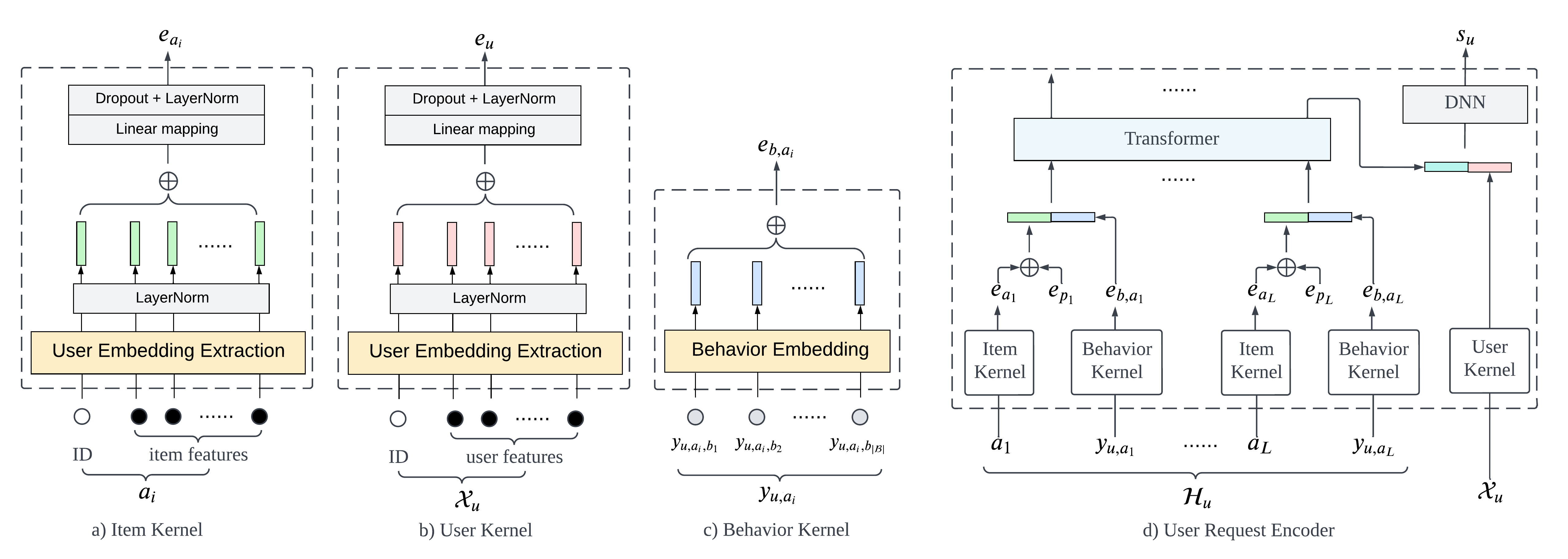}
    \caption{User request encoder module that maps a user request ($\mathcal{X}_u, \mathcal{H}_u$) to a user state vector $s_u$. $e_{a_i}$ and $e_{u}$ represents the item encoding and the user encoding respectively, $e_{p_i}$ represents the learnable positional embedding for items in the history. We set history length $L=50$ in our experiments and use zero encoding for padding item. $\oplus$ represents summation.}
    \label{fig: user_request_encoder}
\end{figure*}

All of our experiments are conducted on a single (Tesla T4) GPU machine with code in PyTorch.
The source code can be found in our GFN4Rec repository https://github.com/CharlieMat/GFN4Rec.

\subsection{Model Specification}\label{app: model_detail}

Figure \ref{fig: user_request_encoder} shows the detail of the user request encoder mentioned in section \ref{sec: user_request_encoder}.
The item and user encoding kernel first aggregates the raw feature embeddings and then apply a linear transformation to give encoding $e_{a_i}$ for item $a_i$ and $e_u$ for user request $u$.
The user response kernel directly aggregates the response embedding of different feedback types without transformation as shown in Figure \ref{fig: user_request_encoder}.
Then the most recent $L$ history interactions $(a_t, y_{u,a_t})$ form a sequence and we use a transformer to generate the history encoding.
Finally, the history encoding and the encoded user profile are mapped to the state encoding $s_u$.
Note that the item kernels are also used in the forward probability estimation in GFN4Rec as shown in Figure \ref{fig: gfn_model}.
In order to keep the model comparison reasonable, we keep the profile and history encoder the same across all models.

\subsection{Sensitivity Analysis:}

During the search of hyper-parameters, in almost all experiments, we found $b\_z=1.0$ consistently gives good results so it is fixed when searching $b\_f$ and $b\_r$.
We then adopted an iterative line search strategy for $\{b_f, b_r\}$:
First, we fix $b_r$ and search for the best $b_f$; then fix $b_f$ with the current best setting and search for $b_r$; we repeat these steps for two rounds and it converges.
As an example, 
Table \ref{tab: gfn_tb_bf_sensitivity} and Table \ref{tab: gfn_tb_br_sensitivity} presents this process of GFN4REC on KR1K.
The resulting best setting for reward metrics is 
$b_r=0.1, b_f=1.0$ for GFN4REC.
ML1M also converges in two rounds but with a slightly different best point 
$b_r=0.3, b_f=0.5$ for GFN4REC.
In our observation, we found that GFN4REC is not so sensitive to b\_f 
near the observed best setting in KR1K, but we kept skeptical of whether this is a universal pattern across different datasets.

\subsection{Computational Complexity}

For a list size K, GFN4Rec will run the forward function K times, and each time it samples an item according to the item probability.
In comparison, the whole list generation method like ListCVAE only runs the forward function 1 time but still needs to do item sampling from a candidate set for K times.
The CF-based methods in our implementation do not engage exploration so have no sampling process, which induces the lowest computational cost.
In our experiments, we use GPUs to process the computation, so the differences in the running time caused by forward function complexity differences become smaller. 
For example, ListCVAE has a total inference+training time of around 4800 seconds for 5000 steps while GFN4Rec uses around 5500 seconds on KR1K.
CF uses 3341 seconds which is the fastest since there is no sampling process.
In general, the overall inference efficiency of GFN4Rec is identical to all existing auto-regressive solutions, and it is mainly controlled by the list size $K$ ignoring the differences in model complexity.

\begin{table}[t]
    \centering
    \begin{tabular}{c|ccccc}
    \hline
$b_f$ & 0.1 & 0.3 & 0.5 & 0.7 & 0.9\\
\hline
Avg $\mathcal{R}$ & 2.379 & 2.359 & 2.268 & 2.371 & 2.367\\
Max $\mathcal{R}$ & 4.083 & 4.066 & 4.050 & 4.058 & 4.042\\
Coverage & 18.976 & 22.161 & 49.588 & 16.894 & 16.800\\
ILD & 0.554 & 0.525 & 0.532 & 0.565 & 0.531\\
\hline
    \end{tabular}
    \caption{Search of $b_f$ for GFN4REC with $b_r=0.5, b_z=1.0$. The best performance is not statistically significant.}
    \label{tab: gfn_tb_bf_sensitivity}
\end{table}

\begin{table}[t]
    \centering
    \begin{tabular}{c|ccccc}
    \hline
$b_r$ & \text{0.1} & 0.3 & 0.5 & 0.7 & 0.9\\
\hline
Avg $\mathcal{R}$ & 2.414 & 2.401 & 2.374 & 2.384 & 2.377\\
Max $\mathcal{R}$ & 4.054 & 4.053 & 4.040 & 4.042 & 4.048\\
Coverage & 21.267 & 19.082 & 18.839 & 18.212 & 18.760\\
ILD & 0.520 & 0.522 & 0.540 & 0.523 & 0.522\\
\hline
    \end{tabular}
    \caption{Search of $b_r$ for GFN4REC with $b_f=1.0, b_z=1.0$.}
    \label{tab: gfn_tb_br_sensitivity}
\end{table}

\end{document}